\documentclass[prb,twocolumn,superscriptaddress,showpacs,%
preprintnumbers,amsmath,amssymb]{revtex4}
\usepackage{amsmath}
\usepackage{graphicx}
\usepackage{dcolumn}

\begin{document}

\title{Theoretical investigation of electron-hole complexes in anisotropic two-dimensional materials}

\author{A. Chaves} \email{andrey@fisica.ufc.br}
\affiliation{Departamento de F\'isica, Universidade Federal do
Cear\'a, Caixa Postal 6030, Campus do Pici, 60455-900 Fortaleza,
Cear\'a, Brazil}
\affiliation{Department of Chemistry, Columbia University, New York, NY 10027, USA}
\author{M. Z. Mayers}
\affiliation{Department of Chemistry, Columbia University, New York, NY 10027, USA}
\author{F. M. Peeters}
\affiliation{Department of Physics, University of
Antwerp, Groenenborgerlaan 171, B-2020 Antwerp,
Belgium}
\affiliation{Departamento de F\'isica, Universidade Federal do
Cear\'a, Caixa Postal 6030, Campus do Pici, 60455-900 Fortaleza,
Cear\'a, Brazil}
\author{D. R. Reichman}
\affiliation{Department of Chemistry, Columbia University, New York, NY 10027, USA}

\date{ \today }

\begin{abstract}
Trions and biexcitons in anisotropic two-dimensional materials are investigated within an effective mass theory. Explicit results are obtained for phosphorene and arsenene, materials that share features such as a direct quasi-particle gap and anisotropic conduction and valence bands. Trions are predicted to have remarkably high binding energies and an elongated electron-hole structure with a preference for alignment along the armchair direction, where the effective masses are lower. We find that biexciton binding energies are also notably large, especially for monolayer phosphorene, where they are found to be twice as large as those for typical monolayer transition metal dichalcogenides.
\end{abstract}


\maketitle

Recently discovered few-layer semiconductors, such as transition metal dichalcogenides (TMDCs) and black phosphorus \cite{Xia, GomezReview} (phosphorene, or bP), have given rise to a class of novel quasi-two-dimensional (2D) materials. Phosphorene, in particular, has attracted interest due to its unique properties, namely a direct band-gap with a highly anisotropic band structure \cite{Katsnelson, Cakir} and high carrier mobilities. \cite{Xia, GomezReview} Technological applications, such as phosphorene-based field effect transistors \cite{Ye} and photodetector devices, \cite{Youngblood, Avouris} have been proposed, suggesting that this material might play an important role in opto-electronics in the near future. Few-layer arsenic (arsenene) has been theoretically demonstrated to be stable and to share many of the electronic properties with phosphorene. In particular, it is also a single-element crystalline layer, with highly anisotropic bands and a direct band gap. \cite{Kamal, ParametersAs} Most importantly, the band anisotropy in both materials gives rise to unique physical properties, e.g. direction-dependent effective masses, \cite{ParametersbP} mobilities, \cite{Xia} and excitonic polarizabilities, \cite{Chaves1} features expected to yield avenues for new device applications. \cite{Millie}

One of the hallmarks of atomically thin two-dimensional materials is their unprecedented high exciton binding energies, which arise as a consequence of the reduced screening of the electron-hole Coulomb interaction by the environment surrounding the layer. Such strong electron-hole interactions allow for the observation of a clear Rydberg series of excitonic states, \cite{Alexey1, Alexey2} as well as resolved trion and biexciton peaks in opto-electronic experiments. \cite{HeinzBiexciton, Experiment, Mak, Singh} In addition, reduced dimensionality leads to deviations from the standard 3D form of the electron-hole interaction, which results in clear deviations from the hydrogenic form of excitonic level spacing. \cite{Keldysh, Cudazzo, Berkelbach, CastroNeto}

Interest in higher order excitonic complexes in black phosphorus has recently been boosted by experimental claims of trion binding energies as high as $\sim$100 meV as measured by photoluminescence. \cite{Experiment} Such a huge trion binding energy would be of great importance, allowing for the possibility that a tightly bound trion could be driven by in plane electric fields without dissociation. This experimental result was rationalized within the context of a quasi-one-dimensional charge-complex model, \cite{Experiment, Thilagam} arising from the high effective mass anisotropy in this system, and highlights the possibility that material anisotropy may lead to novel effects associated with larger electron-hole complexes.

In this work, we calculate trion and biexciton binding energies in anisotropic 2D materials, with special emphasis on $n$-layer black phosphorus ($n$-bP) and the recently theoretically proposed $n$-layer arsenene ($n$-As). Our results show that, although larger than the usual trion and biexciton binding energies in the TMDCs, the trion binding energies in these materials are significantly lower than those claimed in recent experimental works. Possible reasons for this (apparent) discrepancy between experimental and theoretical values are addressed, and the validity of the 1D model \cite{Thilagam} of trion binding energies in this system is questioned.

Within the parabolic approximation for the energy bands of BP, the trion Hamiltonian may be written as
\begin{equation} \label{eq.exc_ham}
H_\pm =  \sum_{i = e,h} \left[\frac{p_{(i)x}^2}{2m^x_{i}} + \frac{p_{(i)y}^2}{2m^y_{i}} \right] + \frac{p_{(\pm)x}^2}{2m^x_{\pm}} + \frac{p_{(\pm)y}^2}{2m^y_{\pm}} + V_{\pm}
\end{equation}
where $m^{x(y)}_{i}$ is the electron ($i = e$) or hole $(i = h)$ anisotropic effective mass in the x-($y$-)direction. The plus (minus) sign stands for a positively (negatively) charged trion, where an extra hole $h'$ (electron $e'$) with effective mass $m^{x(y)}_{+} = m^{x(y)}_{h}$ ($m^{x(y)}_{-} = m^{x(y)}_{e}$) is added through the third and fourth kinetic energy terms to the excitonic pair represented by the first two terms. The interaction potential between charges is assumed to be of the Keldysh form, \cite{Keldysh, Cudazzo, Berkelbach, CastroNeto} with the dielectric screening of the different media above (vacuum) and below (substrate) the semiconductor layer taken into account:
\begin{eqnarray}\label{eq.pot}
V_- =  V_{eh}(\vec r_h - \vec r_{e}) + V_{eh}(\vec r_h - \vec r_{e'}) + V_{ee}(\vec r_{e} - \vec r_{e'}), \nonumber \\
V_+ =  V_{eh}(\vec r_{h} - \vec r_{e}) + V_{eh}(\vec r_{h'} - \vec r_{e}) + V_{hh}(\vec r_{h} - \vec r_{h'}), \nonumber \\
V_{ij} =  \frac {q_iq_je^2}{2(\epsilon_1 + \epsilon_2)\rho_0}\left[H_0\left(\frac{|\vec r_i - \vec{r_j}|}{\rho_0}\right) - Y_0\left(\frac{|\vec r_i - \vec{r_j}|}{\rho_0}\right)\right].
\end{eqnarray}
Here, $q_{e(h)}$ is the electron (hole) charge and $\epsilon_{1(2)}$ is the effective dielectric constant of the vacuum (substrate) above (below) the layer. $H_0$ and $Y_0$ are the Struve and Neumann functions, respectively, and $\rho_0 = n d \epsilon/(\epsilon_1 + \epsilon_2)$ is the screening length, where $d$ is the layer width, assumed to be approximately equal to the inter-layer distance ($5.19$\AA\, and 5.00 \AA\, for $n$-bP and $n$-As, respectively). The effective dielectric constant $\epsilon$ within the layer is taken to be $\epsilon = 10\epsilon_0$ for both materials. \cite{Gomez, footnote0}

Without any further simplification, one can obtain ground state properties of the full Hamiltonian [Eq. (\ref{eq.exc_ham})] by diffusion Monte Carlo (DMC) calculations, \cite{Ceperley, QMCReview} as explained in detail in the Supplementary Material. Since DMC does not require storing a fine mesh and large arrays associated with the wave functions, this method requires much lower computational memory as compared to finite difference-based methods, which makes it suitable for problems with a large number of variables, allowing for the description of larger exciton complexes such as biexcitons. On the other hand, bare DMC calculations do not directly provide the wave functions, although pair correlation functions can still be extracted by the forward walking method. \cite{FWalking}

As an alternative approach, we propose another method for obtaining excitons and trions eigenstates, that is based on the so-called split-operator technique. \cite{AndreySplitOp} Just like DMC, this method uses imaginary time ($\tau = it$) evolution, but now with the imaginary time evolution operator $U(\tau+\Delta \tau,\tau) = \exp(-H\Delta \tau/\hbar)$, whose potential and kinetic energy terms are conveniently split into a series of exponentials, 
\begin{equation}\label{eq.timeevop}
U(\tau+\Delta \tau,\tau) \approx e^{-\frac{\Delta \tau}{2\hbar}V}e^{-\frac{\Delta \tau}{\hbar}T_1}e^{-\frac{\Delta \tau}{\hbar}T_2}...e^{-\frac{\Delta \tau}{\hbar}T_N}e^{-\frac{\Delta \tau}{2\hbar}V} 
\end{equation}
where $T_i$ is the kinetic energy term in the $i$-direction (for a system with $N$ dimensions). This approximation, based on the Suzuki-Trotter expansion, has a $O(\Delta \tau^3)$ error, which is controlled here by using a small imaginary time-step $\Delta \tau$. \cite{AndreySplitOp, Suzuki1, Suzuki2} By propagating an arbitrary initial wave function in imaginary time, this method directly yields the trion ground state energy $E_T$ and the trion wave function, but requires storing a high-dimensional numerical array. In this case, reducing the number of variables of the system is of essence. For negatively charged trions, this is achieved by following the procedure proposed in Refs. [\onlinecite{Thilagam, Xie}], which is generalized for the anisotropic case in the Supplementary Material. This procedure reduces the six-dimensional Hamiltonian, Eq. (\ref{eq.exc_ham}), to a four-dimensional one using the relative coordinates between the first electron and the hole, $\vec{R} = \vec{r}_h - \vec{r}_{e}$, i.e. the exciton coordinates, and between the second electron and the exciton center-of-mass, $\vec{r} = \vec{r}_{e'} - \vec{R}^{cm}_{exc}$. Adapting this procedure to deal with a positively charged trion ($H_+$) is straightforward.  Throughout this paper, we will use either DMC or split-operator methods, whenever convenient. Material parameters for $n$-bP and $n$-As are obtained from Refs. \onlinecite{ParametersbP} and \onlinecite{ParametersAs}, respectively.

\begin{figure}[!b]
\centerline{\includegraphics[width=\linewidth]{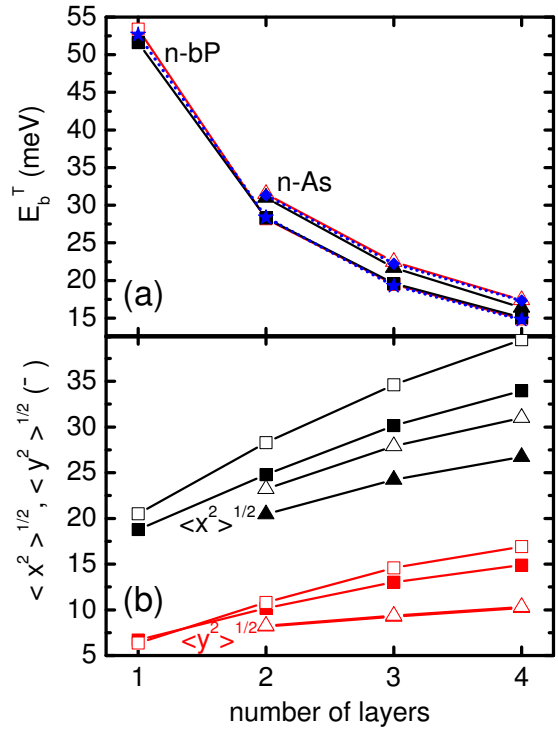}}
\caption{(Color online) (a) Binding energies of positively (red open symbols) and negatively (black full symbols) charged trions in $n$-bP (squares) and $n$-As (triangles), as calculated by DMC and by the split-operator technique (blue stars and diamonds, respectively). DMC calculations have an energy variance $\approx\pm0.3$ meV. (b) Square root of the expectation values of $x^2$ (black) and $y^2$ (red) of the exciton (full symbols) and e-c.m. (open symbols) contributions to the total wave function, for a negatively charged trion in $n$-bP (squares) and $n$-As (triangles), as obtained by the split-operator technique.} 
\label{fig.energies}
\end{figure}

The ground state exciton binding energy $E^X_b$ is obtained by the same methods proposed here, but using the neutral exciton Hamiltonian [see, e.g. Eq. (3) of Ref. [\onlinecite{Chaves1}]]. We then subtract the numerically obtained exciton and trion energies to calculate the trion binding energy $E^T_{b} = E_b^X - E_T$. Figure \ref{fig.energies}(a) shows the positive ($X^{+}$, open red symbols) and negative ($X^-$, closed black symbols) trion energies of $n$-bP (squares) and $n$-As (triangles), obtained by DMC calculations, as a function of the number of layers $n$. We first take a suspended sample, by setting $\epsilon_2 = \epsilon_0$. \cite{footnote}  Since electrons and holes have similar masses in both materials, positive and negative trions have almost the same energy, although a few meV separation can be observed for lower number of layers in $n$-bP, where the electron-hole mass disparity is enhanced. \cite{ParametersbP, ParametersAs} Due to such high similarity between $X^-$ and $X^+$ states, results for the latter are omitted in the subsequent discussion. Results obtained with the split-operator method for $X^-$ are shown for $n$-bP and $n$-As as blue stars and diamonds, respectively, and agree well with the DMC results. 

Charge carriers in these materials exhibit different effective masses along different directions, which results in a distinct spatial extent of the $x$ and $y$ components of the position vectors $\vec{r}$ and $\vec{R}$ defining the trion. These quantities are calculated separately for the exciton ($\vec{R}$) and electron-exciton center-of-mass (e-c.m., $\vec{r}$) contributions to the overall wave function, respectively, which are illustrated in Fig. \ref{fig.energies}(b) for $X^-$ in $n$-bP (squares) and $n$-As (triangles). We observe that both the exciton (full symbols) and the e-c.m. (open symbols) distributions are wider in the (armchair) $x$-direction, due to the lower effective masses in this direction. The anisotropy of the probability distributions in $n$-As is strongly enhanced as the number of layers increases, as a consequence of the increase of the hole effective mass in the $y$-direction. \cite{ParametersAs}

\begin{figure}[!b]
\centerline{\includegraphics[width=\linewidth]{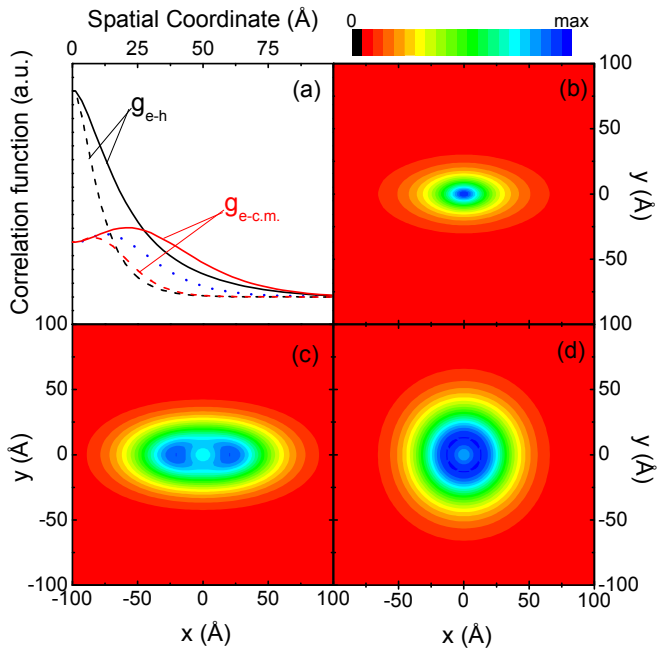}}
\caption{(Color online) (a) Correlation functions for the electron-hole ($g_{e\mbox{-}h}$, black) pair and electron-exciton center-of-mass ($g_{e\mbox{-}c.m.}$, red) at $x = 0$ (solid) and $y = 0$ (dashed) in $1$-bP over a SiO$_2$ substrate. The blue dotted line illustrates $g_{e\mbox{-}c.m.}$ for the isotropic mass case. The full correlation function distributions over the $(x,y)$-plane are shown for (b) $g_{e\mbox{-}h}(x,y)$, (c) $g_{e\mbox{-}c.m.}(x,y)$, and (d) $g_{e\mbox{-}c.m.}(x,y)$ for the isotropic case.} 
\label{fig.correlation}
\end{figure}

The behavior of the electron and hole distributions in anisotropic trions is further illuminated by looking at the correlation functions between the particles. Due to the particular choice of coordinates in our calculations, it is convenient to define
\begin{equation}
g_{e\mbox{-}h}(\vec{\rho}) = \left\langle \delta\left(\vec{\rho} - \vec{R}\right)\right\rangle = \int_{S_r} |\Psi(\vec{r},\vec{\rho})| d^2r
\end{equation}
as the correlation function between electron and hole and
\begin{equation}
g_{e\mbox{-}c.m.}(\vec{\rho}) = \left\langle \delta\left(\vec{\rho} - \vec{r}\right)\right\rangle = \int_{S_R} |\Psi(\vec{\rho},\vec{R})| d^2R
\end{equation}
as the correlation function between the second electron and the exciton center-of-mass. These correlation functions are shown along the $x = 0$ (solid) and $y = 0$ (dashed) axis in Fig. \ref{fig.correlation}(a). The maximum of the e-c.m. function $g_{e-c.m.}$ is clearly displaced from the $(x,y) = (0,0)$ point, which is reasonable, since this function describes the probability distribution for an electron interacting with an exciton, which provides both repulsive (due to the second electron) and attractive (due to the hole) contributions to the potential. Results for $g_{e\mbox{-}c.m.}$ in a system with the same parameters as 1-bP, but assuming isotropic effective masses $m_e^x = m_e^y = 0.17 m_0$ and $m_h^x = m_h^y = 0.15 m_0$, are shown as a blue dotted curve, for comparison. 

The anisotropy of the excitonic component $g_{e\mbox{-}h}$ becomes even more evident in the two-dimensional contour plot of the correlation functions, as illustrated in Fig. \ref{fig.correlation}(b), which is clearly stretched out along the $x$-axis. As for the e-c.m. component, Fig. \ref{fig.correlation}(c) shows that the effective mass anisotropy leads to a function with two clear peaks along the $x$-axis. In fact, since the electrons in the $X^-$ trion repel each other, they are expected to seek the largest possible separation. On the other hand, electrons and holes attract each other. Thus the lowest energy situation that satisfies a large (short) electron-electron(hole) separation is expected to be the one where the three particles form an electron-hole-electron (e-h-e) line. \cite{Thilagam} The e-h-e line could be along any direction in an isotropic material, as verified by $g_{e\mbox{-}c.m.}$ for the isotropic case in Fig. \ref{fig.correlation}(d), where the correlation function is circularly symmetric with a peak at $r \neq 0$. However, since $n$-bP has anisotropic effective masses, the line along the $x$ (armchair) direction, where effective masses are lower, is the most energetically favourable. This linear trion structure has been used \cite{Experiment} in an effectively 1D trion model, following the steps of Ref. [\onlinecite{Thilagam}], from which a trion energy of $\sim 100$ meV was obtained. Among the major issues for the validity of this model in the present context is the fact that it has been developed for a material with isotropic bands, where electrons and holes interact via a bare Coulomb potential, which is not the case for excitons and trions in few-layer phosphorene, as previously discussed. 

\begin{figure}[!t]
\centerline{\includegraphics[width=\linewidth]{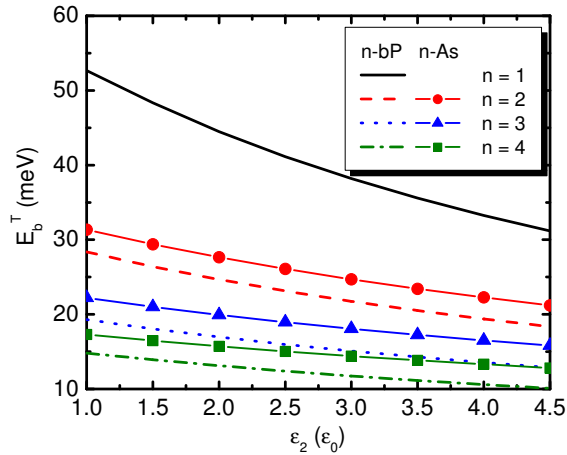}}
\caption{(Color online) Trion binding energies of $n$-bP (lines) and $n$-As (symbols) as a function of the dielectric constant of the substrate.} 
\label{fig.EvsEpsilon}
\end{figure}

So far, the trion binding energies have been obtained for a suspended sample, i.e. without a substrate. Trions in suspended 1-bP are found to have a remarkably high binding energy $E_b^T = 51.6$ meV, as compared to monolayer TMDCs [c.f. Fig. \ref{fig.energies}(a)]. However, in recent experimental measurements of excitons and trion states in $1$-bP, \cite{Experiment, Gomez} the phospherene layer has been deposited on a SiO$_2$ substrate ($\epsilon_2 = 3.8 \epsilon_0$). The trion binding energy in 2D materials is reduced by screening of the electron-hole interaction potential due to its surrounding media. In order to address this issue, we have calculated the dependence of the trion binding energies on the dielectric constant of the substrate, $\epsilon_2$, which is shown in Fig. \ref{fig.EvsEpsilon} for $n$-bP and $n$-As. \cite{footnote} We observe that if one assumes a SiO$_2$ substrate below the layer, the trion binding energy of $1$-bP decreases to $\sim$34.2 meV, which is much lower than the 100 meV claimed in recent experiments. \cite{Experiment} One could argue that, even in the presence of a substrate, the remarkably high 51 meV trion binding energy found for $\epsilon_2 = \epsilon_0$ could still be valid if, for instance, the measurement is made in a region of the layer where it is accidentally suspended from the surface of the substrate. On the other hand, the exciton peak in that experiment was located near 1.75 eV, whereas the exciton binding energy for $\epsilon_2 = \epsilon_0$ in our model is found to be $E^X_b \approx$ 0.744 eV, which implies an electronic (quasi-particle) gap of $\approx 2.494$ eV, inconsistent with \textit{ab initio} band structure calculations available in the literature for this material. \cite{VyTran, Katsnelson, ParametersbP, Gomez} Assuming $\epsilon_2 = 3.8 \epsilon_0$, one obtains $E^X_b \approx 0.405$ eV, suggesting an electronic gap of $\approx 2.155$ eV, which is in good agreement with \textit{ab initio} results and corroborates with the picture of a screened trion, with a lower binding energy of $E_b^T =$ 34.16 meV.

\begin{table}
\caption{Exciton ($E^X_b$) and biexciton ($E^{XX}_b$) binding energies (in meV) obtained by DMC calculations. $E^{XX}$ is the full biexciton energy, i.e. before subtraction of the exciton contribution.} \label{tab:2}\renewcommand{\arraystretch}{1.5}
\begin{ruledtabular}
\begin{tabular}{cccc}
  & $E^X_b$ & $E^{XX}$ & $E_b^{XX}$    \\\hline
 1-bP & 743.9 & 1528.7 &  40.9  \\
 2-bP &  441.8 & 903.4 & 19.8 \\
 3-bP & 318.9 & 650.9 & 13.1  \\
 4-bP & 253.5 & 516.5 & 9.5 \\\hline
 2-As & 497.1 & 1016.1 & 21.9 \\
 3-As & 363.7 & 743.2 & 15.8 \\
 4-As & 291 & 592.8 & 10.8\\
\end{tabular}
\end{ruledtabular}
\end{table}

Finally, we have performed DMC calculations to obtain the biexciton energies $E^{XX}$ in suspended $n$-bP and $n$-As layers, as shown in Table \ref{tab:2}. The biexciton Hamiltonian is obtained from Eq. (\ref{eq.exc_ham}) by simply adding an extra charge and  properly adapting the potential. The biexciton binding energies are obtained as $E_b^{XX} = E^{XX}-2E^X_b$. The exciton binding energies $E^X_b$ are also given in Table \ref{tab:2}, showing fairly good agreement with results from more sophisticated GW-Bethe-Salpeter calculations for $n$-bP. \cite{VyTran2} The biexciton energies are surprisingly large, especially for 1-bP, where they are almost twice as large as those of monolayer TMDCs, \cite{Matt} which should motivate the search for an experimental observation of biexciton peaks in these 2D materials.

In summary, we have calculated trion and biexciton energies in two examples of anisotropic 2D materials: few layer phosphorene and the recently theoretically proposed arsenene. Although binding energies of these charge complexes are found to be much higher than in their bulk counterparts and even higher than in other 2D materials, such as TMDCs, they are much lower than the remarkably high $\sim$100 meV trion energy claimed for phosphorene in recent experiments. \cite{Experiment} Such a high trion binding energy was explained as due to the quasi-1D nature of trions in this system, as a result of its highly anisotropic effective masses. In this case, arsenene would also exhibit such high trion energies, which is not predicted by our theory. It is important to note that the observation of trion binding energies in optical experiments such as photoluminescence and absorption strongly depends on the doping of the sample under investigation. \cite{Alexey} This issue was indeed addressed by the authors of Ref. [\onlinecite{Experiment}], who noted that the excitation power in their photoluminescence experiments is enhanced to reduce the influence of doping on the results. However, this study was performed only within the range from 0.19 to 1.15 $\mu$W, since higher excitation power would damage the sample. Within this range, the trion energy peak monotonically decreases almost linearly, making it difficult to properly estimate the exact converged value of the trion binding energy. Consequently, the 100 meV estimate in this experiment can only be seen as an upper bound, as emphasized by the authors of Ref. [\onlinecite{Experiment}]. In this sense, the lower trion binding energies found in our study, and confirmed by different calculation methods, suggest that either: (i) trions in this system are not properly described by Wannier-Mott theory, or (ii) the upper bound for the trion binding energy observed in recent experiments with bP is still somewhat far from its actual value. Since Wannier-Mott theory has been successfully used to model experimentally obtained excitonic spectra for different 2D materials, and because our model yields exciton binding energies in bP in good agreement with ab initio calculations, it would appear that the Wannier-Mott approach is sound. Our results then imply that more detailed experiments on trions in few layer bP are needed, e.g. investigating the influence of an external gate potential on the trion peak position, so as to reduce the effect of doping by controlling the Fermi level of the system. \cite{Alexey}

\acknowledgments This work has been financially supported by CNPq, through the PRONEX/FUNCAP and Science Without Borders programs, and the FWO-CNPq bilateral program between Brazil and Flanders. MZM is supported  by a fellowship from the National Science  Foundation, under grant number DGE-11-44155.


\begin{references}
\bibitem{footnote0} Notice that, although slightly anisotropic, the dielectric constant of bulk bP in each direction has similar values, $\epsilon_x = 12.5\epsilon_0$, $\epsilon_y = 10\epsilon_0$, and $\epsilon_z = 8\epsilon_0$. Therefore, here we use a geometric mean for the dielectric constant of $n$-bP, $\epsilon = \sqrt[3]{\epsilon_x \epsilon_y \epsilon_z} = 10\epsilon_0$, following Ref. [\onlinecite{Gomez}]. As for $n$-As, dielectric constants are still missing in the literature, thus, we consider here the same value of $\epsilon$ as in bP, which is expected to be at least in the same order of magnitude.
\end{references}
\end{document}

\title{Theoretical investigation of electron-hole complexes in anisotropic two-dimensional materials: Supplementary Material}

\author{A. Chaves} \email{andrey@fisica.ufc.br}
\affiliation{Departamento de F\'isica, Universidade Federal do
Cear\'a, Caixa Postal 6030, Campus do Pici, 60455-900 Fortaleza,
Cear\'a, Brazil}
\affiliation{Department of Chemistry, Columbia University, New York, NY 10027, USA}
\author{M. Z. Mayers}
\affiliation{Department of Chemistry, Columbia University, New York, NY 10027, USA}
\author{F. M. Peeters}
\affiliation{Department of Physics, University of
Antwerp, Groenenborgerlaan 171, B-2020 Antwerp,
Belgium}
\affiliation{Departamento de F\'isica, Universidade Federal do
Cear\'a, Caixa Postal 6030, Campus do Pici, 60455-900 Fortaleza,
Cear\'a, Brazil}
\author{D. R. Reichman}
\affiliation{Department of Chemistry, Columbia University, New York, NY 10027, USA}

\maketitle

\section{Material Parameters}

\begin{table}[!h]
\caption{Electron and hole effective masses for $n$-BP and $n$-As used in our calculations. \cite{ParametersAs, ParametersbP}} \label{tab:2}\renewcommand{\arraystretch}{1.5}
\begin{ruledtabular}
\begin{tabular}{ccccc}
  & $m_e^x$ & $m_h^x$ & $m_e^y$ & $m_h^y$ \\\hline
 1-bP & 0.17 & 0.15 & 1.12 & 6.35  \\
 2-bP & 0.18 & 0.15 & 1.13 & 1.81  \\
 3-bP & 0.16 & 0.15 & 1.15 & 1.12 \\
 4-bP & 0.16 & 0.14 & 1.16 & 0.97 \\\hline
 2-As & 0.25 & 0.21 & 1.38 & 4.49  \\
 3-As & 0.23 & 0.20 & 1.39 & 7.54   \\
 4-As & 0.22 & 0.20 & 1.40 & 10.08 \\
\end{tabular}
\end{ruledtabular}
\end{table}

\section{Split-Operator method: Relative and center-of-mass coordinates}

\begin{figure}[!t]
\centerline{\includegraphics[width=0.9\linewidth]{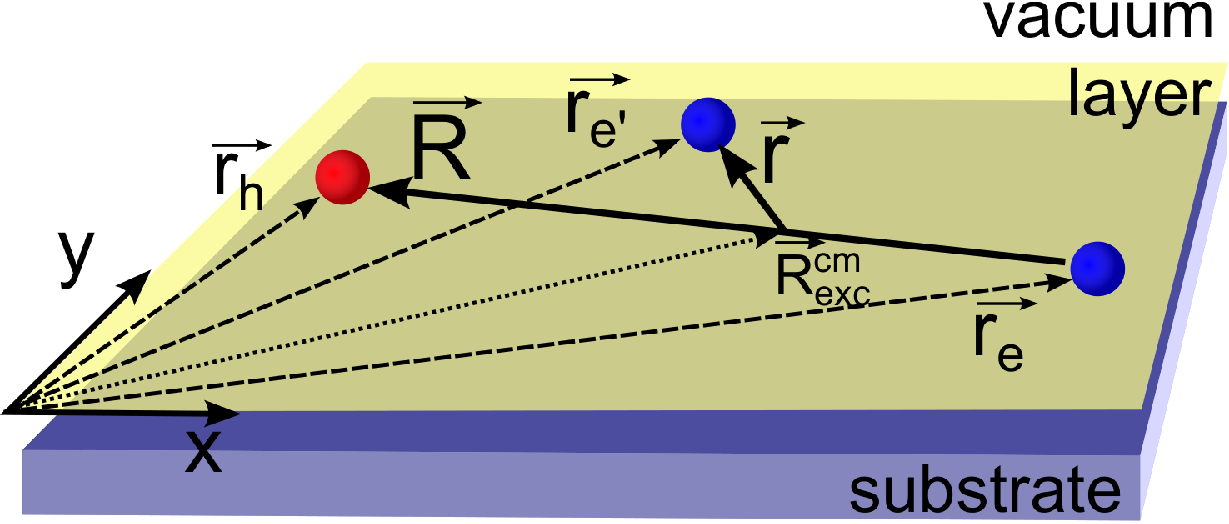}}
\caption{(Color online) Sketch of the exciton center-of-mass ($\vec R^{cm}_{exc}$) and trion relative ($\vec{r}$) coordinates system for a negatively charged trion, formed by a hole (at $\vec r_h$) and two electrons (at $\vec r_{e}$ and $\vec r_{e'}$), in a 2D layer deposited over a substrate.} 
\label{fig.sketch}
\end{figure}

In order to reduce the number of dimensions in our trion calculations with the split-operator technique, we use the relative and center-of-mass coordinates between the first electron and the hole, $\vec{R} = \vec{r}_h - \vec{r}_{e}$ and $\vec{R}^{cm}_{exc}$, with corresponding momenta $\vec P$ and $\vec {P}^{cm}_{exc}$, respectively, to obtain
\begin{equation} \label{eq.HamIntermediate}
H_- =  \frac{P^{cm~2}_{exc,x}}{2\bar{M}^x} + \frac{P^{cm~2}_{exc,y}}{2\bar{M}^y} + \frac{P_{x}^2}{2{\mu}^x} + \frac{P_{y}^2}{2{\mu}^y} + \frac{p_{(e')x}^2}{2m_e^x} + \frac{p_{(e')y}^2}{2m_e^y} + V_{-},
\end{equation}
where $\bar{M}^{x(y)} = m_e^{x(y)} + m_h^{x(y)} $ and ${\mu}^{x(y)} = (1/m_e^{x(y)} + 1/m_h^{x(y)})^{-1}$ are the anisotropic total and reduced effective masses of the first electron-hole pair, respectively. We then employ a similar transformation in Eq. (\ref{eq.HamIntermediate}), but now using the second electron (related to the last two kinetic energy terms) and the center-of-mass between the first pair (related to the first two kinetic energy terms) \cite{Thilagam}: $\vec{r} = \vec{r}_{e'} - \vec{R}^{cm}_{exc}$, $\vec{R}_{cm} = \vec{R}^{cm}_{exc} + (m^x_ex_{e'}\hat{x} + m_e^y y_{e'}\hat{y})/M$, with corresponding momenta $\vec{p}$ and $\vec{P}_{(cm)}$, respectively, leading to
\begin{equation} \label{eq.HamIntermediate2}
H_- =  \frac{{P}_{(cm)x}^2}{2{M}^x} + \frac{{P}_{(cm)y}^2}{2{M}^y} + \frac{P_{x}^2}{2{\mu}^x} + \frac{P_{y}^2}{2{\mu}^y} + \frac{p_{x}^2}{2\mu_t^x} + \frac{p_{y}^2}{2\mu_t^y} + V_{-},
\end{equation}
where $M^{x(y)}$ is the total anisotropic effective mass of the three particles system in the $x(y)$-direction. Finally, since 
\begin{eqnarray}
\vec r_h - \vec r_{e} = \vec{R} \quad\quad\quad\quad\quad\quad\quad\quad\quad\quad\quad\quad\quad\quad\quad\quad \nonumber \\
x_{e'} - x_{h} = x - \frac{m^{x}_e}{\bar{M}^{x}}X, \quad\quad y_{e'} - y_{h} = y - \frac{m^{y}_e}{\bar{M}^{y}}Y \nonumber \\
x_{e'} - x_{e} = x + \frac{m^{x}_h}{\bar{M}^{x}}X, \quad\quad y_{e'} - y_{e} = y + \frac{m^{y}_h}{\bar{M}^{y}}Y
\end{eqnarray}
the potential term $V_- (|\vec{r}_h - \vec{r}_e|, |\vec{r}_{e'} - \vec{r}_h|,|\vec{r}_{e'} - \vec{r}_e|) = V_- (\vec{r}, \vec{R})$ does not depend on the total center-of-mass coordinate $\vec R_{cm}$ of the system, allowing us to take the first two terms of $H_-$ in Eq. (\ref{eq.HamIntermediate2}) to be zero and, thus, reduce our problem to (numerically) solving a (non-separable) 4-dimensional Schr\"odinger equation, which is feasible with modern computers. This choice of coordinates is schematically represented in Fig. \ref{fig.sketch} for a negatively charged trion, i.e. with two electrons and one hole, represented by blue and red dots, respectively. Adapting this procedure to deal with a positively charged trion ($H_+$) is straightforward. 

\section{Diffusion Monte Carlo: Jastrow term}

Since the diffusion Monte Carlo technique has been widely used and frequently reviewed in the literature, we will restrict ourselves here to the discussion of the details relevant for our model. For a more complete account of the diffusion Monte Carlo (DMC) technique, we refer to e.g. Ref. [\onlinecite{Needs}] and the Supplementary Material of Ref. [\onlinecite{Matt}]. As in the split-operator technique, DMC calculations consist of solving the imaginary time Schr\"odinger equation to project out excited states from an arbitrary initial wave function $\Psi(\vec{r},t=0)$, so that after a long imaginary time propagation, the wave function converges to the ground state of the system. However, in this approach, one defines a mixed probability $f(\vec{r},t) = \Psi_T(\vec{R})\Psi(\vec{r},t)$, which leads to an equation of motion for $f(\vec{r},t)$ that can be solved by approximate Greens functions that result in drift-diffusion motion and branching action.\cite{Needs} The guiding function is defined as $\Psi_T(\vec{r}) = e^{J(\vec{r})}$, with a Jastrow term that, in this work, contains two-body electron-hole and electron-electron contributions adapted specifically for the case of negative trions in the Keldysh potential [Eq. (2) of the main manuscript], \cite{Aleiner}
\begin{equation}
J(\vec{r}) = u_{ee}(\tilde{r}_{ee'}) + u_{eh}(\tilde{r}_{eh}) + u_{eh}(\tilde{r}_{e'h}),
\end{equation}
where
\begin{equation}
u_{eh} (r) = c_1 r^2\ln(r)e^{-c_2r^2} - c_3 r (1-e^{-c_2r^2})
\end{equation}
and
\begin{equation}
u_{ee} (r) = c_4 r^2 \ln(r) e^{-c_5r^2}.
\end{equation}
with $c_1 = 1\big/4r_0$, $c_4 = -1\big/4r_0$, and other parameters optimized as to improve the efficiency of the Monte Carlo sampling. These equations are straightforwardly adapted to describe positively charged trions, just by replacing $u_{ee} \rightarrow u_{hh}$, $\tilde{r}_{ee'} \rightarrow \tilde{r}_{hh'}$, and $\tilde{r}_{e'h} \rightarrow \tilde{r}_{eh'}$. The modified mass-dependent electron-hole separation
\begin{eqnarray}
\tilde{r}_{ij} = \left[\left(\sqrt{m^{x}_{i}}x_{i} - \sqrt{m^{x}_{j}}x_{j}\right)^2 + \right. \quad\quad\quad\quad\quad\nonumber \\
\left. \quad \quad\quad\quad \left(\sqrt{m^{y}_{i}}y_{i} - \sqrt{m^{y}_{j}}y_{j}\right)^2\right]^{1/2},
\end{eqnarray}
with $i$, $j$ $= e$, $h$, $e'$ or $h'$, properly accounts for the anisotropic effective mass of black phosphorus.